\documentclass{icarus}
\usepackage[centertags]{amsmath}
\usepackage{wasysym}
\usepackage{icarus}

\newcommand{\e}[2]{$#1~\cdot~10^{#2}$}

\renewcommand{\baselinestretch}{1}
\journame{The journal of Planetary and Space Science} \pages{13}
\tables{1} \figures{0} \email{bonnie@pmfst.hr} \submitted{Jan 17
2006} \revised{July 18 2006}

\begin{document}

\title{On the low-mass planethood criterion}
\author{Bojan Pe\v{c}nik}
\author{Christopher Broeg}


\section*{Abstract}
We propose a quantitative concept for the lower planetary boundary,
requiring that a planet must keep its atmosphere in vacuum. The
solution-set framework of \cite{PecnikWuchterl_05} enabled a clear
and quantitative criterion for the discrimination of a planet and a
minor body. Using a simple isothermal core-envelope model, we apply
the proposed planetary criterion to the large bodies in the Solar
System.

\section{Introduction}
The number of known planets has increased by more than an order of
magnitude within the last decade - for the current list of
extrasolar planets see e.g.\ \cite{SchneiderExoEnc}. Those
additional planets created a diversity which made the task of
defining \emph{what is a planet} all the more difficult. Three
distinct groups of properties are relevant when discussing the
planethood criteria: physical characteristics, planetary dynamics,
and cosmogony \citep[for discussion see e.g.\ ][]{Basri_PlanetDef}.
\paragraph{Orbital dynamics} While the majority of the community agrees
that dynamically dominant bodies bound to stars or stellar remnants
should be counted as planets (provided that they're not massive
enough to support fusion), presently there's no consensus on how to
designate potential unbound planetary-mass objects. Disagreement
partly stems from a lack of understanding of the formation
process(es) that could create unbound planetary-mass objects. The
current guideline of the International Astronomical Union's (IAU)
Working Group on Extrasolar Planets \citep[][]{wgesp_definition}
classifies such bodies not as `planets', but as `sub-brown dwarfs'.
\paragraph{Cosmogony} The present knowledge on the
cosmogony of planets in general \citep[e.g.][ for an
overview]{ppiv_book} is far from being complete. On the other hand,
for a foreseeable time there will be no way to observationally infer
the history of exoplanets in detail. It is difficult to use
cosmogony for any general planethood criterion which would be
determinable from easily observed characteristics \citep[e.g.\
][]{Basri_PlanetDef, SternLevinson}. Hence, WGESP guideline
\citep[][]{wgesp_definition} does not include cosmogony in the
planet definition.
\paragraph{Physical properties} The `planetary mass' has its upper boundary around
13 M$_{\mathrm\jupiter}$ (M$_{\mathrm\jupiter}$ being Jupiter mass)
as a deuterium burning limit \citep[e.g][]{wgesp_definition,
Basri_PlanetDef, SternLevinson}, but also as an effective upper end
of the extrasolar planet mass-distribution
\citep[][]{marcy_butler_00, SchneiderDB}. The observational limit
suggests that 13 M$_{\mathrm\jupiter}$ effective boundary is a
consequence of an as-of-yet not fully understood formation process,
and that the deuterium burning limit is either a side effect or a
numerical coincidence. The emerging 1 M$_{\mathrm\jupiter}$
characteristic mass \citep{SchneiderDB} also questions the
importance of the thermonuclear reactions as the planetary
delimiter. We should note that the recent trend of discovering
Neptune-class planets \citep[e.g.][]{Butler04, Bonfils05} could
further reduce the observed characteristic mass.

Various concepts have been proposed for the lower planetary mass
limit, but the official consensus is not yet reached. Concepts are
based either on a particular physical property of the body in
question, or on setting the arbitrary value for the required mass or
radius of the body. An arbitrary boundary is based either on a
historical heritage (e.\,g.\ discovery of Pluto) or on some
numerical peculiarity of the Solar System (e.\,g.\ size of Ceres).
In the past decade, the diversity of members of planetary family has
been shown to grossly exceed examples from the Solar System. Hence,
we feel that the general definition of a planet cannot include
arbitrary values introduced from our own planetary neighborhood.\\*
Continuing on the physical concepts for the low-mass planetary
boundary, a few similar ideas are based on roundness of the body,
i.\,e. a body must be massive enough for its self-gravity to
dominate over any material forces that might produce asymmetric
shapes \citep[e.g.][]{Basri_PlanetDef, SternLevinson}. The minimum
size of such body somewhat depends on its composition, but one can
calculate it to be a bit less than 500 km in diameter
\citep[][]{Basri_PlanetDef}. However, much smaller bodies can also
achieve round shape via melting, or other form of asteroid
differentiation; for review see \cite{McCoy04, Taylor93}.
\cite{McCoy04} write that already a 10 km iron meteorite could reach
90\% partial melting early in the formation process, and `Given that
iron meteorite parent bodies were in the range of 10-100 km in
radius, high degrees of partial melting must have been reasonably
widespread, given the abundance of iron meteorite groups and
ungrouped meteorites.' Additionally, from the round shape of
km-sized rubble pile asteroids \citep[e.g.\ Dactyl, moon of Ida,
c.f.][]{Davis96} we know that roundness can be acquired via violent
kinetic events.\\* These facts hint that `roundness' could be
connected to the cosmogony of the object rather than to its physical
properties only, and as such would not be an ideally suited
observable. Hence, roundness is not a proper concept at the bottom
of the planetary mass scale.

The presence of other physical properties, such as a magnetic field,
an atmosphere, or the existence of a satellite, has been proposed as
the planethood criterion. \citet{SternLevinson} show that any single
such criterion is unable even to include all of the Solar system
planets.

While the concept for the lower planetary mass limit is not yet a
big problem for the extrasolar-planet surveys, it is the subject of
a heated debate concerning the Solar system: Pluto's double
classification (both as a minor planet and a planet) and the
upcoming issue of what to do with its potentially even larger
cousins from the trans-neptunian-object family
\citep[e.g.][]{Brown05}. It has been suggested
\citep[e.g.][]{marcy_butler_00} -- completely arbitrary and mostly
for historic reasons -- to accept Pluto's mass as the lower mass
delimiter. Others have argued the same for the mass of Ceres
\citep[e.g.][]{Basri_PlanetDef}. IAU has no official position yet,
with WGESP opting for `The minimum mass/size required for an
extrasolar object to be considered a planet should be the same as
that used in our Solar System.'

In the next section we present our version of the low-mass planetary
boundary. The third section summarizes the model of
\cite{PecnikWuchterl_05}, which we use in the fourth section to
apply our planethood criterion to the large bodies in the Solar
system. The last section summarizes the paper with conclusions and
remarks.

\section{Low-mass planetary boundary}
\label{Sect_Def_Planet}

During the course of our investigation, we have developed a concept
for a global static critical core mass (\cite{PecnikWuchterl_05}, to
which we refer further on as Paper I). We make use of this concept
to provide a planethood criterion to discriminate a planet from a
lesser body (eg. an asteroid). Alternatively, it could be a boundary
between a `major' and a `minor' planet. The following discourse
should be valid in addition to fulfilling the dynamical planethood
criteria, i.e.\ being the gravitationally dominant body in an orbit
around a star or a stellar remnant.

As previously stated, the upper mass limit for a giant planet has to
be below the limiting mass for thermonuclear fusion of deuterium
\citep{wgesp_definition}. The lower mass limit for a giant planet is
often assumed without much deliberation, but could be stated as: `An
object is a giant planet if it is able to keep its atmosphere,
composed mainly from the solar-composition gas of the parent body,
against the surrounding vacuum.'

In the absence of the well-quantifiable physical property which
would distinguish a rock (minor planet) from a (terrestrial) planet,
we propose to apply the planethood criterion valid for giant planets
also to the terrestrials. Previous similar concepts required a body
to have an atmosphere. This was unsatisfactory because this
requirement excluded atmosphereless planets (e.g. Mercury),
\emph{and} had a difficulty in setting the limit between an
`atmosphere' and a `vacuum'.

We argue that an object should not be called a \emph{planet} if it
is not \emph{capable} to retain its envelope (volatiles) when
connected to vacuum (i.e.\ to an empty space, as opposed to the
proto-planetary nebula gas cloud). We do not require an atmosphere,
just the capability to retain it. The survey of the hydrostatic
proto-planetary core-envelope systems (Paper I) describes the
envelope properties, depending on the protoplanetary core mass
relative to the critical core mass ($M_\mathrm{crit}$). Our concept
of the global static critical core mass enables us to quantitatively
discriminate a `gaseous envelope' from a `vacuum'. Therefore, one of
the characteristics that a celestial body must fulfill to be
called a planet can be specified as:\\
\emph{A planet will have a core which is supercritical within the
appropriate manifold. A minor planet will have a subcritical core.}
\\
`Manifold' is a complete set of equilibrium states relevant to the
particular planetary environment (see Paper I for detailed
discussion). Name `manifold' is chosen to describe a particular
property of the solution set, where every subcritical core connects
to a particular nebula cloud with multiple (\emph{at least} two)
envelope solutions. A supercritical core, on the other hand, always
connects planet's atmosphere to a vacuum of the surrounding space
with just one, unique solution.

We go further, and define a \emph{`giant planet'} to be a
supercritical core within the respective manifold for a solar
(parent star) gas composition. Obviously, main atmosphere
constituent for a \emph{`telluric planet'} cannot be
solar-composition gas. Several alternatives are available, including
molecular nitrogen, molecular oxygen, carbon-dioxide, or some gas
mixture. In order to be able to make predictions with our
core-envelope model, we chose molecular nitrogen as a main gaseous
constituent, i.e.\ a `telluric planet' should be supercritical
within the respective manifold for a nitrogen atmosphere, thus
satisfying the \emph{terrestrial planethood criterion} (TPC).

We have chosen molecular nitrogen in the case of a telluric planet
for several reasons, but we could have decided for some other gas
with similar properties. Here we briefly present arguments for
nitrogen. First, it is substantially heavier than hydrogen or
helium, which is a key property enabling a body significantly
smaller than e.\,g.\ Saturn's solid core to retain a gas envelope in
the empty space. Second, the only two other bodies having
significant atmosphere in the Sol system (excluding the giant
planets, of course) both have nitrogen in their atmospheres, giving
(partial) atmospheric pressure similar to the one on Earth.
Admittedly, Venus has carbon-dioxide as a main atmosphere
constituent, but that could be due to the ongoing geological
process(es). Besides, the molecular weight of carbon-dioxide is not
very different from the molecular weight of nitrogen and thus would
lead to a similar classification, in the framework of our model.

It is important to note here that our planethood criterion is based
on the ability to discern the envelope properties from the different
regions of the complete planetary solution set (manifold). The
criterion is \emph{not} based on the simple isothermal model
presented in Paper I. Therefore, \emph{the proposed planethood
criterion definition is general, and its validity extends beyond our
isothermal model to any complete protoplanetary solution set,
regardless of the complexity of the physical model used.}

\section{Model}
Here we present the model of \cite{PecnikWuchterl_05}, which is used
in next section to apply the planethood criterion to large bodies in
the Solar system.
\subsection{Model Assumptions}

We approximate the protoplanet as a spherically symmetric,
isothermal, self-gravitating classical ideal gas envelope in
equilibrium around a core of given mass. This gaseous envelope is
that required to fill the gravitational sphere of influence,
approximated by the Hill-sphere:
\begin{equation}\label{r_Hill_gen}
  r_{\mathrm Hill}=a\sqrt[3]{M_{\mathrm planet}/3M_{\star}},
\end{equation}
where $a$ is the orbital distance from a parent star. Having a mean
molecular weight $\mu$ of $1.4~10^{-2} \; \mathrm{kg\,mol^{-1}}$ the
protoplanetary envelope approximates a pure molecular nitrogen
atmosphere. The protoplanet's heavy-element-core is represented by a
rigid sphere of uniform density.

\subsection{Model Equations}

The envelope is set in isothermal hydrostatic equilibrium, with
spherical symmetry, and as such is described by:\\*
\begin{equation}\label{poisson}
  \frac{\textit{d}M(r)}{\textit{d}r}=4\pi r^2\varrho(r),
\end{equation}
the equation of hydrostatic equilibrium:
\begin{equation}\label{hyd_equil}
\frac{\textit{d}P(r)}{\textit{d}r}=-\frac{GM(r)}{r^2}\varrho(r),
\end{equation}
and the equation of state for an ideal gas:
\begin{equation}\label{id_gas}
  P(r)=\frac{\Re T}{\mu}\varrho(r).
\end{equation}
$M(r)$ is defined as the total mass (core plus envelope) contained
within the radius $r$:
\begin{equation}\label{mass_radial}
  M(r)=M_{\mathrm core}+\int_{r_{\mathrm core}}^r4\pi r'^2\varrho(r')\,dr',
\end{equation}
where $r$ is the radial distance measured from the core center and
$\varrho$ is the envelope gas density at radial distance $r$.\\* The
inner and outer radial boundaries are:
\begin{equation}\label{r_core}
  r_{\mathrm in}=r_{\mathrm {core}}=\sqrt[3]{\frac{M_{\mathrm core}}{\frac{4}{3}\pi\varrho_{\mathrm
  core}}} \;\;\;\; {\mathrm{and}} \;\;\; r_{\mathrm out}=r_{\mathrm Hill}.
\end{equation}
An additional boundary condition at the core surface is:
\begin{equation}\label{rho_core}
  \varrho_{\mathrm env}(r_{\mathrm core})=\varrho_{\mathrm cs},
\end{equation}
where $\varrho_{\mathrm cs}$ is the envelope gas density at the
\underline{c}ore \underline{s}urface. This model, together with the
specified assumptions and boundary conditions, is sufficient to
completely determine a single model-protoplanet. The total mass and
nebula density at $r_{\mathrm Hill}$ (gas density at protoplanet's
outer boundary) are results of the calculation. The envelope mass
integration is performed using an ADA95 code implementing a
Bulrisch-Stoer-Richardson extrapolation scheme. The total
protoplanetary mass is obtained by integrating outward from
$r_{\mathrm core}$ to $r_{\mathrm Hill}(M_{\mathrm tot})$, using
$\varrho_{\mathrm cs}$ as the starting value for the envelope
density.

\section{Planethood
affiliation in the Solar system} \label{SectPlanetAffiliation}
%

In order to demonstrate a real-world application of the proposed
planethood criterion, we used the model presented in a previous
Section to calculate solution sets (manifolds) with orbital and
environmental characteristics of all known large bodies in the Solar
System. Furthermore, from those manifolds we extracted values for
the global critical core masses ($M_\mathrm{crit}$). Next, we used
those $M_\mathrm{crit}$ values as lower limits for the quantitative
(i.e.\ `realistic') planethood criterion, and compared them to the
real masses of investigated objects. Results are summarized in
Table~1. In the next subsection we discuss the limitations of the
used isothermal model.

From Table~1 we see that terrestrial planets indeed satisfy the TPC,
except maybe Pluto (and perhaps even Mercury) which is very close to
the lower limit of the terrestrial planethood criterion (TPC). A
more sophisticated planet model is necessary to conclude if Pluto
indeed satisfies the criterion.

Besides analyzing manifolds of the known planets, we have also
looked into the planethood criteria for the (larger) moons, asking:
`If the present-day moon's primary were a star \citep[e.g.\
see][]{Clarke68}, what role would the orbiting body assume?' \\*
Some giant planets' moons (Io, Titan, Ganymede, Europa) satisfy the
TPC even better than some planets (Pluto, Mercury, Mars). Of course,
we do not propose to classify those moons as \emph{planets}, but in
our opinion it would be useful to somehow distinguish moons that
satisfy the TPC from those that do not.

\subsection{Limits of the isothermal model}
\label{SectApplicability}

The quantitatively correct $M_\mathrm{crit}$ cannot be determined
through the isothermal model. We can presently only estimate the
range for the realistic $M_\mathrm{crit}$, using the isothermal
model as the lower limit, and the real planet with an atmosphere as
the upper limit.\\*
A relevant isothermal manifold is determined with the appropriate
values for: an orbital distance from the primary, a mass of the
primary, an envelope gas temperature, a mean molecular weight (of
the envelope gas), and a solid core density. See \cite{pecnik05} and
Paper I for the model details, for the discussion on the solution
manifold, and for detailed analysis of the applicability of the used
isothermal model.

While the values of most of the model parameters needed to calculate
manifolds relevant for large Sol system bodies were straightforward
\citep[e.g.\ see][]{NASASolSysExpl, SolSystParam}, the choice for
the manifold temperature is not completely trivial (see Paper I,
Sect.\ 3.10 for the discussion of the mass-temperature scaling of
the manifold). For our purpose we use the present-day surface
temperature of the investigated bodies \citep[e.g.\
see][]{NASASolSysExpl}. A present-day surface temperature is
relatively good (mass weighted) approximation of the average
atmosphere temperature of the Solar system's terrestrial planets and
Titan. However, such temperatures are much lower that the ones
expected close to core-surface of the just-critical planet.
Additionally, a realistic-gas pressure is larger than the ideal-gas
pressure at high gas densities, because of the finite-volume effects
of the individual gas particles. Choice of the present-day surface
temperature and ideal gas, along with the isothermality of the
model, is appropriate for a subcritical core with relatively low gas
density at the core surface, but will underestimate the
gravitational potential necessary to keep the envelope gas pressure
in the hydrostatic balance for the just-critical model. In other
words, the isothermal $M_\mathrm{crit}$ is smaller than the
$M_\mathrm{crit}$ of a manifold which includes detailed
micro-physics. Hence, our \emph{isothermal $M_\mathrm{crit}$ can
only be used as a lower limit for the quantitative TPC estimate.}

The empirical upper limit for the realistic $M_\mathrm{crit}$ can be
the mass of an existing large body with a significant atmosphere.
From Table~1, we see that such a (supercritical) body (e.g.\ Mars)
is about 40 times more massive than the isothermal
$M_\mathrm{crit}$. More stringent constraint would be possible if
there would be a celestial body in the Solar system which is less
supercritical than Mars (i.e.\ has a smaller ratio of total to
critical core mass), and still has a significant (detectable)
atmosphere. Since the Solar system is not rich enough to place
stronger constraint on the isothermal model, we presently can not
conclude if the atmosphereless bodies, which are less than 40 times
more massive than the respective isothermal $M_\mathrm{crit}$,
satisfy the qualitative TPC (e.g.\ we cannot classify Pluto using
the simple isothermal model).\\* It should be noted that if a
supercritical object (e.g.\ Mercury) does not have an envelope, but
at the same time it could have one according to our classification,
there is no contradiction. This simply means that significant
amounts of nitrogen gas (or any other gas with a relatively high
mean molecular weight) were not available during the
formation/evolution of the body, or have been lost. The body itself
would still qualify as a planet, on the condition that its core is
supercritical, i.e.\ only the \emph{capacity to retain} the
atmosphere is required, not the atmosphere itself.

\section{Summary and Discussion}
\paragraph{Summary} We provide a new lower planet boundary concept, requiring that a
planet must be able to keep an atmosphere in vacuum. The previous
attempts to use the atmosphere as a planetary delimiter had problems
differentiating a dilute atmosphere from a vacuum. With the manifold
paradigm \citep[][]{PecnikWuchterl_05} we completely avoid the issue
of weak atmospheres, and use the critical core mass, a crucial
quantity of the planet formation, to discriminate a planet from a
minor body. Because of such a general framework, the criterion
stands a good chance to be relevant both for planets within and
outside of the Solar system.

The proposed planethood concept is not sensitive to individual
cosmogony, and requires only observables which are available from a
transit event (mass, density, orbital distance) or can be extracted
from the observations of the primary (primary mass, temperature).
Only rough planetary classification can be performed with the
current isothermal model. However, the proposed planethood concept
will be quantitatively more accurate with a better physical
description of a planet \citep[for an example of a manifold with
better constitutive relations see][]{Broeg05}.

\paragraph{Outlook} The extrasolar planet surveys are presently yielding hot Neptunes -
10-15 $M_\mathrm{\earth}$ close-in planets. The advances in the
observational technology and methods will further increase the
sensitivity, and the trend of discovering ever-smaller planets
\citep[see e.g. Fig.\ 1 in][]{santos05} is likely to continue. The
upcoming wide-field transit surveys (e.g. COROT, Kepler) will extend
the range of the known extrasolar planets to the Earth-mass regime.
A clear, easily determinable, and quantitative concept for the lower
planetary mass is urgently needed.

\paragraph{Arbitrary planethood} Before the intricate process(es) of the planet formation and
evolution are sufficiently understood, \emph{every planethood
concept}, including the one we champion, \emph{will be completely
arbitrary}. While being arbitrary, it should reflect in the best
possible way what the astronomical community or maybe even humanity
as a whole considers 'a planet'. We think that the ability to retain
an atmosphere could be such a criterion. At the same time this
criterion is broad enough that we believe it to be applicable to all
the expected and unexpected planet discoveries in the future.

We discussed present ideas on the lower mass limit. We showed that
roundness of the celestial body can be interwoven with cosmogony,
and as such is not a good observable. Additionally, if the cosmogony
would be accepted as the part of the planet definition, it would be
impossible (in the foreseeable future) to decide if most of the
exoplanetary candidates are indeed planets. This would be quite
unpractical. Once we can infer details of the history of exoplanets
through observation, \emph{and} once we have in-depth knowledge of
the general theory of planet formation, cosmogony \emph{might}
become useful component of the planet definition. On the other hand,
an arbitrary fixed value for the object's mass as the limit (e.g.
Pluto) might be irrelevant outside of the Solar system.

We base our \emph{arbitrary} concept on a physical property (as
opposed to a historical paradigm, or to a fixed number coming from
Sol system), and, using the Occam's razor, apply the same reasoning
that appears to work well for giant planets.

With the growing knowledge of the (exo)planetary properties, along
with a healthy debate in peer-refereed publications, we think that
the astronomical community stands a good chance of agreeing on a
proper planethood criteria in the forseable future.

\textbf{Acknowledgement} \\*We thank G\"{u}nther Wuchterl for the
valuable discussions. His comments greatly improved this paper. We
also thank anonymous referees whose comments helped to make this
manuscript clearer. This research was in part supported by
Max-Planck-Institut f\"{u}r extraterrestrische Physik, by National
Foundation for Science, Higher Education and Technological
Development of the Republic of Croatia, and by DLR project number
50-OW-0501.

\def\baselinestretch{1.0}
   \begin{table}[p]
         \label{Tab_Sol_Sys_CCM}
     $$
         \begin{tabular}{l|c|c|c|c|c|c|c}
            \hline
            \noalign{\smallskip}
            object & $M_\mathrm{primary}$ & orb. dist.& $\varrho_\mathrm{core}$& $T_\mathrm{cs}$/[K]& $M_\mathrm{crit}$/[kg]&$M_\mathrm{obj}/M_\mathrm{crit}$&TPC\\
                   &                      &   $a$/[AU]&          /[kg m$^{-3}$]&                    &                                   &                                &satisfied     \\
            \noalign{\smallskip}
            \hline
            \hline
            \noalign{\smallskip}
            Mercury & $M_\mathrm{\odot}$ & 0.387 & 5427 & 440 & \e{2.51}{22} & 13.2  & \checkmark ?\\
            Venus   & $M_\mathrm{\odot}$ & 0.723 & 5240 & 737 & \e{7.08}{22} & 68.7  & \checkmark \\
            Earth   & $M_\mathrm{\odot}$ & 1.0   & 5515 & 287 & \e{1.91}{22} & 312.8 & \checkmark \\
            Mars    & $M_\mathrm{\odot}$ & 1.523 & 3934 & 210 & \e{1.58}{22} & 40.6  & \checkmark \\
            Jupiter$^{\dagger}$ & $M_\mathrm{\odot}$ & 5.203 & 5515 & 153 & \e{1.12}{22} & - & - \\
            Saturn$^{\dagger}$  & $M_\mathrm{\odot}$ & 9.537 & 5515 & 143 & \e{1.17}{22} & - & - \\
            Uranus$^{\dagger}$  & $M_\mathrm{\odot}$ & 19.19 & 5515 &  68 & \e{4.47}{21} & - & - \\
            Neptune$^{\dagger}$ & $M_\mathrm{\odot}$ & 30.07 & 5515 &  53 & \e{3.16}{21} & - & - \\
            Pluto   & $M_\mathrm{\odot}$ & 39.48 & 2000 &  44 & \e{4.05}{21} & 2.5 & \checkmark ? \\
            Ceres   & $M_\mathrm{\odot}$ & 2.767 & 2050 & 167 & \e{1.66}{22} & 0.057 & $\times$ \\
            Moon    & $M_\mathrm{\oplus}$&0.00254& 3344 & 250 & \e{1.0}{22} & 7.35 & \checkmark ?\\
            Io & $M_\mathrm{\jupiter}$& 0.0028& 3528 & 130 & \e{1.0}{21}  & 89.32  & \checkmark \\
            Europa&$M_\mathrm{\jupiter}$&0.004486&3013& 103& \e{1.12}{21} & 42.78 & \checkmark \\
            Ganymede&$M_\mathrm{\jupiter}$&0.00715&1942&115& \e{2.24}{21} & 66.14  & \checkmark \\
            Callisto&$M_\mathrm{\jupiter}$&0.01259&1834&115& \e{3.16}{21} & 34.05 & \checkmark ?\\
            Mimas&$M_\mathrm{\saturn}$&0.00124& 1152 & 70  & \e{1.1}{20} & 0.34 & $\times$ \\
            Enceladus&$M_\mathrm{\saturn}$&0.00159&1606&70 & \e{2.5}{20} & 0.28 & $\times$ \\
            Tethys&$M_\mathrm{\saturn}$&0.00197& 956 & 86  & \e{4.47}{20} & 1.40 & \checkmark ? \\
            Dione &$M_\mathrm{\saturn}$&0.00252& 1500 & 87  & \e{7.08}{20} & 1.55 & \checkmark ? \\
            Rhea  &$M_\mathrm{\saturn}$&0.00352& 1240 & 76  & \e{1.0}{21}  & 2.32 & \checkmark ? \\
            Titan &$M_\mathrm{\saturn}$&0.00817& 1880 & 94  & \e{2.24}{21} & 60.1 & \checkmark \\
            Iapetus&$M_\mathrm{\saturn}$&0.02381& 1088 & 76  & \e{3.55}{21} & 0.45 & $\times$ ?\\
            \noalign{\smallskip}
            \hline
         \end{tabular}
     $$
     \def\baselinestretch{1}
     \caption[]{Terrestrial Planethood Criterion (TPC) for Solar System objects
     calculated with \citet{PecnikWuchterl_05} planetary model.
     Celestial bodies are assumed to fulfill TPC if they can keep an equilibrium nitrogen
     atmosphere of temperature $T_\mathrm{cs}$, in a vacuum, orbiting at
     a distance $a$ from a parent star of mass $M_\mathrm{primary}$.
     Solar system giant planets are
     supercritical even for a solar composition gas (hence also for nitrogen), thus we only
     show the value for the nitrogen-supercritical object, at the
     present-day locations of the giant planets. See text for
     discussion. Symbols in the last column: `\checkmark' - TPC fulfilled, `\checkmark ?' - TPC fulfilled
     for the isothermal \citep{PecnikWuchterl_05} model, unknown if TPC is satisfied for the model with more detailed constitutive
     relations, `$\times$ ?' - TPC most likely not fulfilled even for the isothermal manifold, `$\times$' -
     TPC not fulfilled. \\*
     $^{\mathrm{\dagger}}$ At the locations of gas giants we show a critical core mass for Earth's
     density\\*
     }
   \end{table}
\def\baselinestretch{1.66}

\bibliographystyle{plainnat}
\bibliography{BojanBib}

\end{document}